\documentclass[a4paper,11pt]{article}
\usepackage{amssymb}
\usepackage{graphicx}
\usepackage{amsmath}
\usepackage{bbm}
\begin{document}

\title{
Localization of quantum topology in the presence of matter and gauge fields }
\author{ F. Atyabi \footnote{farzaneh.atyabi@gmail.com} \\
\footnotesize { Department of Physics, Alzahra University, Tehran, Iran} \\}

\maketitle

\newtheorem{The}{Theorem}
\newtheorem{lem}{Lemma}
\newtheorem{prop}{Proposition}
\newtheorem{hyp}{Hypothesis}
\newtheorem{rem}{Remark}
\markright{}

\begin{abstract}
 In this paper a toy model of quantum topology is reviewed to study effects of matter and gauge fields on the topology fluctuations. In the model a collection of $N$ one dimensional manifolds are considered where a set of boundary conditions on states of Hilbert space specifies a set of all topologies perceived by quantum particle and probability of having a specific topology is determined by a partition function over all the topologies in the context of noncommutative spectral geometry. In general the topologies will be fuzzy with the exception of a particular case which is localized by imposing a specific boundary condition. Here fermions and bosons are added to the model. It is shown that in the presence of matter, the fuzziness of topology will be dependent on $N$, however for large $N$ the dependence is removed similar to the case without matter. Also turning on a particular background gauge field, can overcome the fuzziness of topology to reach a localized topology with classical interpretation. It can be seen that for large $N$ more opportunities can be provided for choosing the background gauge field to localize the fuzzy topology.
\end{abstract}

\maketitle

\section{Introduction}

Quantum gravity is one of the most important and challenging issues in theoretical physics. Near Planckian scales, quantum gravitational effects may induce quantum fluctuations in the structure of spacetime.
Over the last few decades, it has been speculated that the stochastic fluctuations of spacetime can emerge as fluctuations of topology in addition to fluctuations in geometrical properties of spacetime, deserving remarkable consideration to achieve a consistent theory of quantum gravity.\\
As we know General Relativity(GR) proposes classical geometric picture of spacetime as a manifold relating the distribution of mass in the universe to its geometry but the topology of the universe is not constrained. However non-manifold structure of spacetime in quantum level may no longer be described by the geometric picture proposed by GR. So a generalization of ordinary geometry for describing quantum structure of spacetime should be searched.\\
One of the most promising candidate to generalize the geometry is Connes' noncommutative geometric construction, which can be expected to provide an appropriate framework to study quantum geometry of the spacetime.
 Therefore to consider both quantum geometry and topology fluctuations of the spacetime, one can use the noncommutative geometric formulation and topological concepts in quantum mechanics.\\
 The idea of changing topology, using properties of wave functions in quantum mechanical systems, was proposed in \cite{Balachandran}.
In Ref\cite{Albuquerque}, topology change of a toy model has been considered by varying boundary condition imposed on quantum states in Hilbert space, where the formalism of spectral action in (non)commutative spectral geometry(NCSG) provides the motivation to define dynamics for the boundary condition.\\
In this paper we intend to consider whether the quantum topology can be affected by coupling with matter and gauge fields and how a localization of the topology can be found as a topology with classical interpretation, which the content is structured as follows: in section~\ref{sec:level1,2}, it is given a summary of the main ingredients of NCSG developed by Connes and collaborators, in which field theories are incorporated by the spectral action principle. In section~\ref{sec:level1,3}, we have a brief review on a toy model of quantum topology fluctuations\cite{Albuquerque}. In sections ~\ref{sec:level1,4} and ~\ref{sec:level1,5} effect of coupling with fermions and gauge bosons on the topology fluctuations is investigated. As it will be explained the topology fluctuations can be controlled by coupling with some particular gauge potentials to get a localized topology.

\section{\label{sec:level1,2}Basic notions of NCSG  }

Within the framework of (non)commutative geometry(NCG), the main properties of a compact spin Riemannian manifold can be recaptured using a spectral triple $( A,H,D)$, where $A$ is a commutative $\mathbb{C}^*$-algebra represented on Hilbert space $H$ (of $L^2-$spinors) by linear operators, and $D$ is an unbounded self-adjoint operator called Dirac operator which is not in $A$ but acts on $H$, satisfying precise conditions \cite{Connes,Landi}.

In the context of NCSG, to obtain the dynamics of a physical theory based on geometry, it is used a universal action functional proposed by Connes and Chamseddine based upon the spectral action principle \cite{Chamseddine,connes2} , i.e. the requirement that the action should only depend on the spectrum of the Dirac operator. The action consists of a bosonic and a fermionic part:
\begin{eqnarray}
S=Tr \chi(D^2/\Lambda^2)+<\psi|D\psi>,
\end{eqnarray}
where $\chi$ in the bosonic part is a suitable cut-off function selecting eigenvalues below the cut-off $\Lambda$, $\psi$ in the fermionic part is the spinor field and $<.|.>$ denotes the inner product on $H$ and the gauge bosons of the theory appear naturally as inner fluctuations of the Dirac operator\cite{Connes,Landi}.

\section{\label{sec:level1,3}Quantum topology fluctuations }
In Ref \cite{Albuquerque}, a toy model for topology fluctuations has been proposed, where boundary condition on states of Hilbert space imposed by some unitary matrices, encodes the topology. The corresponding topology of configuration space perceived by the quantum particle depends on the choice of the unitary matrices such that for some choices, the topologies are localized with classical interpretation however for the others, the topologies can be considered as superposition of the localized ones. So one might think of probability of finding a localized topology which can be computed using partition function. Applying the boundary condition on eigenvectors of Dirac operator, the Dirac spectrum and consequently the spectral action can be written in terms of the eigenvalues of the corresponding unitary matrices and by summing over the set of all the unitary matrices, the partition function can be computed.\\
Let us mention the model a bit more precise. Consider a set  $X$ of $N$ disjoint one dimensional manifolds which are intervals of length $L$. For the collection of one dimensional manifolds, the spectral triple is $(A_X,H_X,D)$, where $A_X$ is the algebra of continuous function on $X$, $H_X$ is $L^2(X)$ and $D$, the Dirac operator, is the momentum operator:
\begin{eqnarray}
D=\left(
                                                                            \begin{array}{cccc}
                                                                             -i\partial_x& 0& \ldots& 0\\
                                                                              0& -i\partial_x& \ddots& \vdots \\
                                                                              \vdots& \ddots& \ddots& 0\\
                                                                              0& \ldots& 0& -i\partial_x
                                                                            \end{array}
                                                                          \right)\nonumber
 \end{eqnarray}
 By considering boundary condition as $\psi(L)=g\psi(0)$ on any state $\psi$ in Hilbert space, parameterized by unitary matrices $g\in U(N)$, we have a fixed spectral triple and spectrum of $D$ may be written as $\frac{2\pi}{L}(n_k+\frac{\alpha_k}{2\pi})$, where $e^{i\alpha_k}$ is the eigenvalue of the matrix $g$ and $k=1,...,N$. The set of all possible $g$'s determines the set of all topologies perceived by quantum particle and a partition function can be defined over all topologies(all the boundary conditions):
 \begin{eqnarray}
 Z=\int [dg] e^ {-S[D(g)]},
  \end{eqnarray}
  where $[dg]$ is U(N)-invariant Haar measure. The action $S$ in the framework of NCSG, only depends on the spectrum of D which are in terms of $\alpha_k$'s. So the partition function can be written as\cite{Mehta}:
 \begin{eqnarray}
Z=\int_{0}^{2\pi}\frac{1}{N!(2\pi)^N}\prod_{k=1}^N d\alpha_k\prod_{i<j}|e^ {i\alpha_i}-e^{i\alpha_j}|^2 e^ {-S(\alpha_l)},
\label{eq:four}
\end{eqnarray}
and probability of finding a topology determined by $g$ can be obtained using the partition function. However the probability distribution depends on the eigenvalues of $g$ and all matrices $g'=ugu^\dag$ $(u\in U(N))$ correspond to the same spectrum and consequently the same probability. Therefore in general the topology is fuzzy with the exception of the case $g=\mathbbm{1}$ which can be considered as a localized topology with classical interpretation\cite{Albuquerque}.

\section{\label{sec:level1,4}Effect of coupling with matter on the topology fluctuations }
Here we intend to investigate whether coupling with matter affects the quantum topology. For non-fermionic part of the spectral action, let us restrict our attention to the one considered in Ref\cite{Albuquerque}, which is Wilson's action for a 2-d Yang-Mills gauge theory on a single plaquette\cite{Gross}:
 \begin{eqnarray}
S(g,\beta)=\frac{N\beta}{2}(g+g^\dag)\label{eq:three},
\end{eqnarray}
and add fermionic part as $<\psi|D\psi>$:
 \begin{eqnarray}
 S(g,\beta)=\frac{N\beta}{2}(g+g^\dag)+<\psi|D\psi>\label{eq:two}.
 \end{eqnarray}
 So partition function will be:
 \begin{eqnarray}
Z=\int d[g]\int \mathcal{D}\overline{\psi}\mathcal{D}\psi   e^{\frac{-N\beta}{2}(g+g^\dag)-<\psi|D(g)\psi>},
\end{eqnarray}
which can be written as:
\begin{eqnarray}
Z= \int_{0}^{2\pi}\frac{1}{N!(2\pi)^N}\prod_{k=1}^N d\alpha_k\prod_{i<j}|e^ {i\alpha_i}-e^{i\alpha_j}|^2e^{-N\beta\sum_k\cos\alpha_k}\int \mathcal{D}\overline{\psi}\mathcal{D}\psi   e^{-<\psi|D\psi>}.
\end{eqnarray}
For grassmann fields $\psi$ and $\overline\psi$:
\begin{eqnarray}
\int \mathcal{D}\overline{\psi}\mathcal{D}\psi   e^{-<\psi|D\psi>}&&=det D\nonumber\\
&&=\prod_{k=1}^{N}\prod_{n_k=-\infty}^{\infty}\frac{2\pi}{L}(n_k+\frac{\alpha_k}{2\pi})\nonumber
\end{eqnarray}
The product over $n_k$ can be written as $C \sin \frac{\alpha_k}{2}$, using the identity $ \prod_{n=1}^{\infty}(1-\frac{x^2}{n^2\pi^2})=\frac{\sin x}{x}$, where $C$ is a divergent coefficient independent of $\alpha_k$.
So we have:
\begin{eqnarray}
Z= \int_{0}^{2\pi}\frac{1}{N!(2\pi)^N}\prod_{k=1}^N d\alpha_k\prod_{i<j}|e^ {i\alpha_i}-e^{i\alpha_j}|^2e^{-N\beta\sum_k\cos\alpha_k}\prod_{k=1}^NC\sin \frac{\alpha_k}{2}
\end{eqnarray}
Applying the following identity for finite $N$ \cite{Forrester}:
\begin{eqnarray}
\frac{1}{N!}\int_{0}^{2\pi}\prod_{k=1}^N d\alpha_k\prod_{l=1}^N f(\alpha_l)\prod_{i<j}|e^ {i\alpha_i}-e^{i\alpha_j}|^2=det[\int_{0}^{2\pi} d\alpha f(\alpha) e^{i\alpha(i-j)}]_{i,j=1,...,N},
\end{eqnarray}
the partition function reduces to:
\begin{eqnarray}
Z=det[C\int_{0}^{2\pi}  d\alpha \sin \frac{\alpha}{2} e^{-N\beta\cos\alpha} e^{i\alpha(i-j)}]_{i,j=1,...,N}
\label{eq:one}
\end{eqnarray}
Although the partition function is ill-defined, the divergence can be eliminated while yields finite expectation values. Here we are interested in obtaining expectation value of $\cos\alpha_k$ to compare with the case without interference of fermions\cite{Albuquerque}, numerically computed in two limits $\beta\rightarrow 0$ and $\beta\rightarrow\infty$  :
\begin{eqnarray}
<\cos\alpha_k>_N(\beta)=-\frac{1}{N^2}\frac{\partial}{\partial\beta}{ln Z_N(\beta)}=\left\{
                                                                            \begin{array}{cc}
                                                                             -\frac{1}{2N+1}& \beta\rightarrow0\\
                                                                              -1& \beta\rightarrow\infty
                                                                            \end{array}
\right.
\end{eqnarray}
For $\beta\rightarrow\infty$, $\alpha_k$'s concentrate around $\pi$ which yields $g\simeq -\mathbbm{1}$. So a localization of topology can be seen by quantum particle, exactly the same as the case not including matter\cite{Albuquerque}.
For $\beta\rightarrow 0$, the expectation value depends on N such that for large N tends to zero while in the case not including matter for any value of $N$, $<\cos\alpha_k>$ tends to zero which results a fuzzy topology\cite{Albuquerque}.

\section{\label{sec:level1,5}Fermions in a fixed gauge background and localization of quantum topology }
In this section, we investigate whether a fuzzy topology can be controlled by turning on a background gauge field.\\Here the fermions are coupled to a gauge background field $A$. In the context of NCSG, gauge bosons appear naturally as inner fluctuations of Dirac operator, which here we consider as:
\begin{eqnarray}
D=D_0+A,\nonumber
\end{eqnarray}
where $D_0$ denotes the momentum operator. Consequently the Dirac spectrum change to $\frac{2\pi}{L}(n_k+\frac{\alpha_k+<A>}{2\pi})$, where $<A>$ is the average of $A(x)$ in an interval of length L.
To compute partition function, let us first consider just non-fermionic part of the spectral action, i.e. Eq.~(\ref{eq:three}), so the partition function Eq.~(\ref{eq:four}) will be:
\begin{eqnarray}
Z= \int_{0}^{2\pi}\frac{1}{N!(2\pi)^N}\prod_{k=1}^N d\alpha_k\prod_{i<j}|e^ {i\alpha_i}-e^{i\alpha_j}|^2e^{-N\beta\sum\cos(\alpha_k+<A>)}
\end{eqnarray}
One can obviously see that partition function is the same as the case without considering gauge field\cite{Albuquerque}. Because the integrand is periodic and with a change of variable $\alpha_k+<A>\rightarrow \alpha_k$, Z does not change. So
\begin{eqnarray}
<\cos\alpha_k>_N(\beta)=\left\{
                                                                            \begin{array}{cc}
                                                                             0& \beta\rightarrow0\\
                                                                              -1& \beta\rightarrow\infty
                                                                            \end{array}
\right.
\end{eqnarray}
 as in Ref\cite{Albuquerque}.\\
Now consider both bosonic and fermionic part of the action. For $\beta\rightarrow\infty$, the fermonic part of the action in Eq.~(\ref{eq:two}), can be neglected, therefore as above we have $<cos\alpha_k>=-1$ so the topology is localized.\\ For finite values of $\beta$ the partition function in Eq.~(\ref{eq:one}) will change to:
\begin{eqnarray}
Z(N,\beta,<A>)=det[C\int_{0}^{2\pi}d\alpha \sin (\frac{\alpha+<A>}{2}) e^{-N\beta\cos(\alpha+<A>)} e^{i\alpha(i-j)}]_{i,j=1,...,N}.
\end{eqnarray}

\begin{figure}
    \centering
\includegraphics[width=90mm, height=40mm]{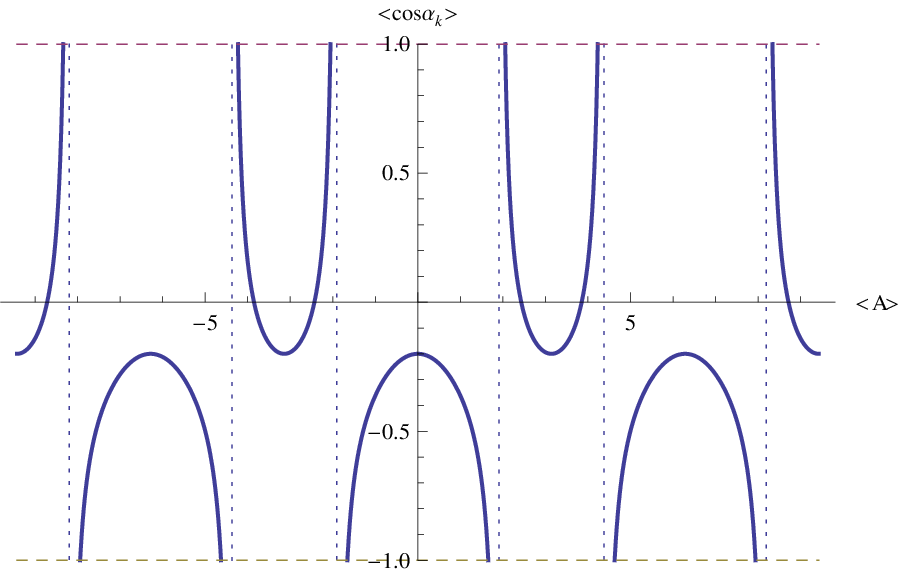}
        \caption{ $N=2$}
       \label{fig 1}
       \end{figure}

\begin{figure}
    \centering

\includegraphics[width=90mm, height=40mm]{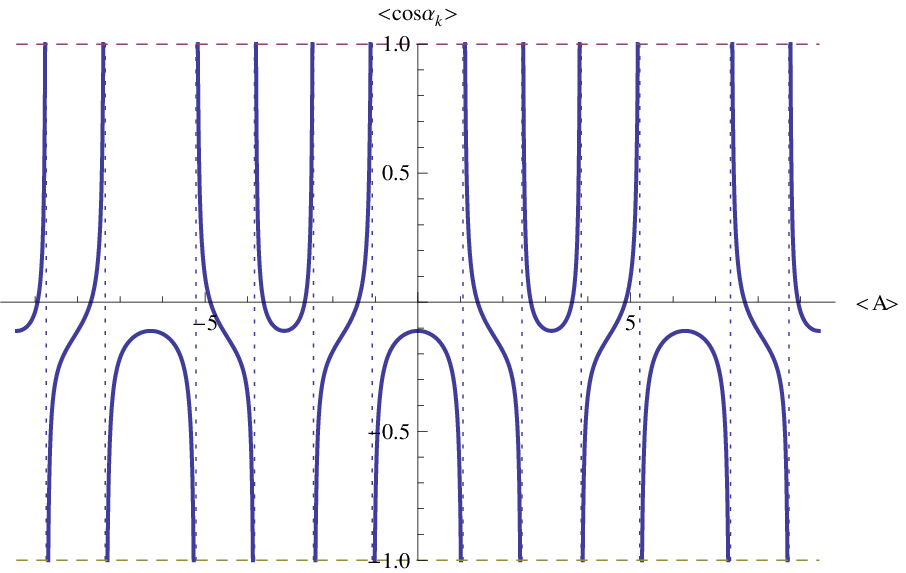}
        \caption{ $N=4$}
       \label{fig 2}
        \end{figure}

\begin{figure}
    \centering

\includegraphics[width=90mm, height=40mm]{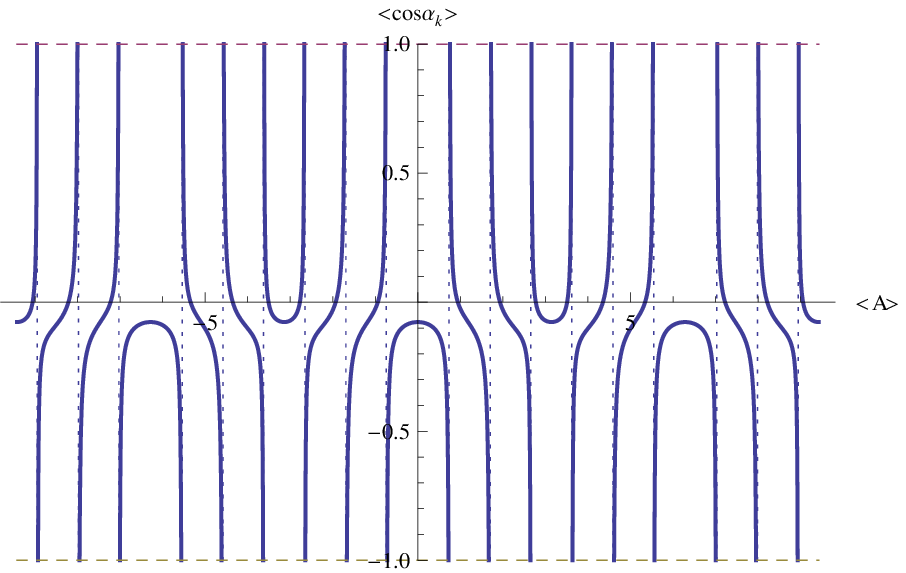}
        \caption{$N=6$}
        \label{fig 3}
        \end{figure}

In the limit $ \beta\rightarrow0$, as an example for $N=2$ we have:

\begin{eqnarray}
\lim_{\beta\rightarrow 0}<\cos\alpha_k>|_
                  {N=2}
=\frac{-8-6 cos<A>+6 cos^2<A>}{10+30 cos<A>}.
\end{eqnarray}
  For some values of $<A>$, $\lim_{\beta\rightarrow 0}<cos\alpha_k>$ will be $\pm1$, and the topology can be localized. In Figs.~\ref{fig 1},~\ref{fig 2} and ~\ref{fig 3}, $<cos\alpha_k>$ against $<A>$ is plotted for $N=2,4,6$. As it can be seen, for larger values of $N$, number of solutions to the equation $<cos\alpha_k>=\pm1$ in a specific range of $<A>$ will be larger. So a particular gauge choice can moderate topology fluctuations to get a localized topology and for large $N$ more opportunity can be provided for choosing the gauge field $A$ which yields a localized topology with classical interpretation.

\section{Conclusion}
We have considered a toy model of fluctuations in topology for a collection of $N$ one dimensional manifolds to study effects of matter and gauge fields on fluctuating topology. The quantum topology is specified by imposing boundary condition on Hilbert states and it's dynamic is determined in the framework of NCSG. As the non-fermionic part of the spectral action, a one plaquette $U(N)$ gauge theory with only one parameter $\beta$ is considered. So the partition function has a single free parameter $\beta$.
It has been shown that in the limit $\beta\rightarrow\infty$, topology is localized similar to the case not including matter and gauge fields. In the limit $\beta\rightarrow 0$, with the presence of matter, fuzziness of topology depends on $N$ and for large $N$ the fuzziness will be similar to the case without matter. We have discussed that in this limit, turning on a particular background gauge field can eliminate quantum fluctuations of topology to accede a localized and classical topology.

\newpage

 \clearpage

\end{document}